# DOING BATTLE with the SUN:
# LESSONS FROM LEO and OPERATING a SATELLITE CONSTELLATION in the ELEVATED ATMOSPHERIC DRAG ENVIRONMENT of SOLAR CYCLE 25

W. Scott Shambaugh

*Capella Space, San Francisco, California, USA, scott.shambaugh@capellaspace.com*

## ABSTRACT

Capella Space, which designs, builds, and operates a constellation of Synthetic Aperture Radar (SAR) Earth-imaging small satellites, faced new challenges with the onset of Solar Cycle 25. By mid-2022, it had become clear that solar activity levels were far exceeding the 2019 prediction published by the National Atmospheric and Oceanic Administration's (NOAA) Space Weather Prediction Center (SWPC). This resulted in the atmospheric density of low Earth orbit (LEO) increasing 2–3x higher than predicted. While this raises difficulties for all satellite operators, Capella's satellites are especially sensitive to aerodynamic drag due to the high surface area of their large deployable radar reflectors. This unpredicted increase in drag threatened premature deorbit and reentry of some of Capella's fleet of spacecraft.

This paper explores Capella's strategic response to this problem at all layers of the satellite lifecycle, examining the engineering challenges and insights gained from adapting an operational constellation to rapidly changing space weather conditions. A key development was the implementation of "low drag mode", which increased on-orbit satellite lifetime by 24% and decreased accumulated momentum from aerodynamic torques by 20–30%. The paper shares operational tradeoffs and lessons from the development, deployment, and validation of this flight mode, offering valuable insights for satellite operators facing similar challenges in LEO's current elevated drag environment.

## 1  BACKGROUND

LEO is not a vacuum. Satellites orbiting between ~80 and ~700 km fly through the thermosphere, an upper layer of Earth's atmosphere where gas floats in free molecular flow conditions. This thin air exerts a drag force that causes a spacecraft's altitude to gradually decay over time, until it hits the denser mesosphere and burns up in atmospheric reentry.

A satellite's acceleration due to drag $\vec{a}_{drag}$, is a function of atmospheric density $\rho$, velocity relative to the air stream $\vec{v}$, coefficient of drag $C_d$, planar area in the velocity direction $A$, and mass of the satellite $m$ (Equation 1). The last terms are specific to the design of a satellite and can be grouped into a single term called the ballistic coefficient $C_B$ (Equation 2).

$$\vec{a}_{drag} = -\frac{1}{2}\rho\vec{v}|\vec{v}|\frac{C_d A}{m} \qquad (1)$$

$$C_B = \frac{C_d A}{m} \qquad (2)$$

While most of these terms are relatively stable, thermosphere density varies widely in both space and time (see Section 4.3). In the 400–600 km range, every 10 km increase in altitude corresponds to a ~14% decrease in density. Density also varies strongly with solar flux and geomagnetic conditions. Solar flux over the 11-year rising and falling solar cycle pumps energy into the thermosphere, causing it to swell and densify at higher altitudes. This solar flux is generally measured by the number of visible sunspots (SSN) or the intensity in the 10.7 cm radio band (F10.7), which are strongly correlated metrics that can be converted between [1]. These nonlinearities in density must be carefully handled when predicting satellite drag.

In most cases, the phenomenon of satellite lifetime is looked at through the lens of minimizing space debris. The Federal Communications Commission (FCC) which issues communication licenses for commercial U.S. satellite operators, requires that spacecraft operating in LEO and licensed under its Part 25 "standard" rules reenter within 25 years after an up-to 15-year operational lifetime. This reentry time drops to five years in October 2024 [2]. Spacecraft licensed under its "streamlined" small satellite rules must have a total in-orbit life of no more than six years [3]. The lifetime analysis must follow NASA guidance in [4] and generally be performed with NASA's Debris Assessment Software (DAS) [5]. From this lens, higher drag environments and rapid post-mission deorbit are preferred, and predictions of a lower density atmosphere make for conservatively long lifetime estimates.

A commercial satellite operator looks at the drag environment from the opposite perspective. For them, the goal is to ensure they meet their minimum design lifetime, and generally to maximize mission duration past that as far as regulatory limits permit. To do this they must protect for higher drag environments that would cause a premature end to on-orbit operations. As Solar Cycle 25's growth towards maximum over the past few years has greatly exceeded predictions, the delta between lower and higher density estimates has taken satellite operators by surprise, often with dramatic consequences.

In February 2022, SpaceX lost 38 of 49 Starlink satellites during a geomagnetic storm immediately after launch to a 340 x 210 km staging orbit, due to atmospheric drag increasing up to 50% higher than expected [6]. This event brought attention to the need for better "nowcast" space weather predictions so that satellite operators can anticipate and react to short term space weather events [7], however such improvements to real time data would not help satellite designers with their need to predict the trajectory of the solar cycle many years into the future.

In August 2023, a TechCrunch news article shared that Capella Space's satellites were having their own struggles with the stronger than predicted solar cycle [8]. The company's Whitney-generation SAR satellites were deorbiting faster than expected and reentering sooner than their intended three-year lifetime. This paper shares a behind-the-scenes look at Capella's thinking and response to this issue.

## 2 INTRODUCTION to CAPELLA SPACE

### 2.1 Capella Space and its Mission

Capella Space is an American space company which designs, manufactures, and operates a constellation of Synthetic Aperture Radar (SAR) Earth-imaging small satellites. Founded in 2016, it was the first U.S. company to fly a constellation of SAR satellites and continues to deliver the highest quality SAR imagery commercially available anywhere in the world. Using X-band radar allows for any-time imaging of anywhere on Earth, regardless of cloud cover or lighting conditions. This represents a significant advantage in imaging availability over optical sensors, which are prevented by clouds and darkness of night from resolving ~75% of Earth's surface at any given time (and are fully denied from extreme latitudes during polar night). This technology is key to the company's mission to make timely Earth observation an essential tool for commerce, conservation, and well-being. Figure 1 shows an example of Capella's imagery.

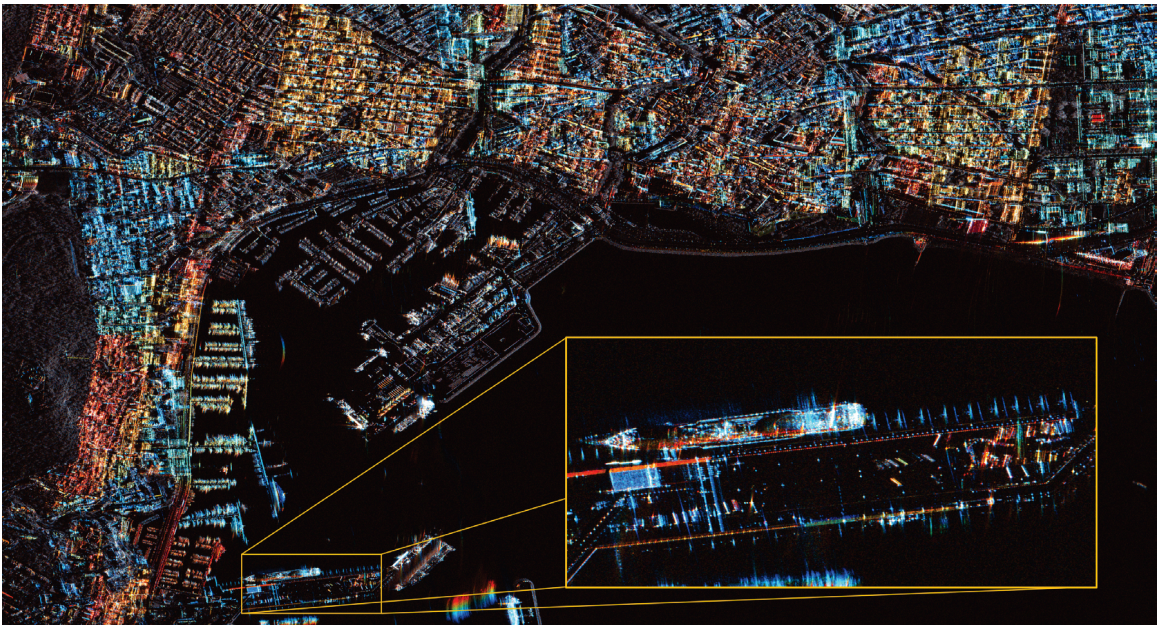

Figure 1: A spotlight SAR image of Palma de Mallorca, Spain taken in June 2022 by the first Whitney-generation satellite, Capella-3. Inset shows details of a ship at port. Colors indicate the squint angle over the duration of the collect, and brightness indicates intensity. See [9] for details.

Figure 2 shows the timeline of Capella's satellite launches. This paper will focus on the six Whitney-generation satellites which were in operation in 2022.

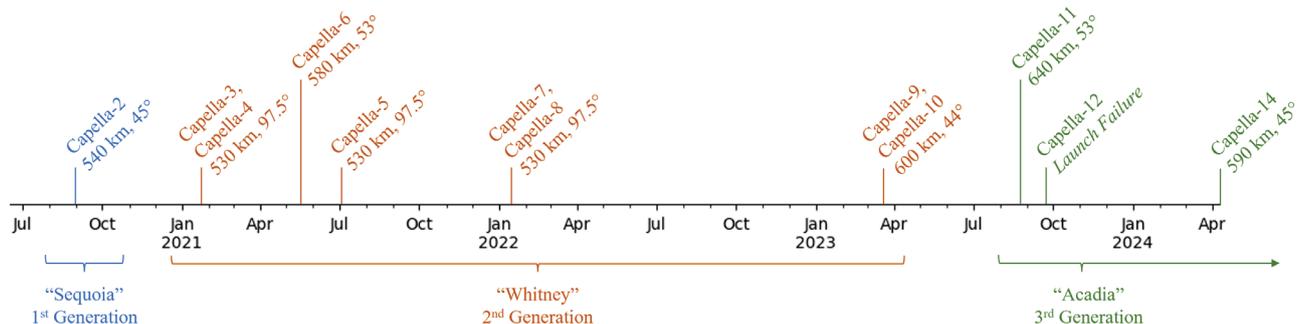

Figure 2: Capella's launch timeline, with satellite name, generation, target altitude, and inclination.

## 2.2 The Capella "Whitney" Satellite

The Whitney satellite design is distinguished by several large deployable structures which unfold from the satellite bus after the satellite is delivered to orbit. These deployable structures consist of solar panels, a 3-meter folding boom which carries the radar instrument, and a 3.5-meter diameter radar reflector made from fine wire mesh. To capture images, the radar instrument emits microwave pulses that bounce off the reflector towards an Earth target. The return echo is concentrated by the same reflector back to the radar instrument, and this signal data is downlinked over the satellite's high gain antenna and then processed to generate the SAR images. Whitney also carried an antenna to communicate with Inmarsat's constellation of geostationary communication satellites via their Inter-satellite Data Relay Service (IDRS), which allows the satellite to share telemetry and be tasked to take imagery in near real time without needing to wait to fly over a ground station. See Figure 3 for a rendering of the satellite with these components labeled.

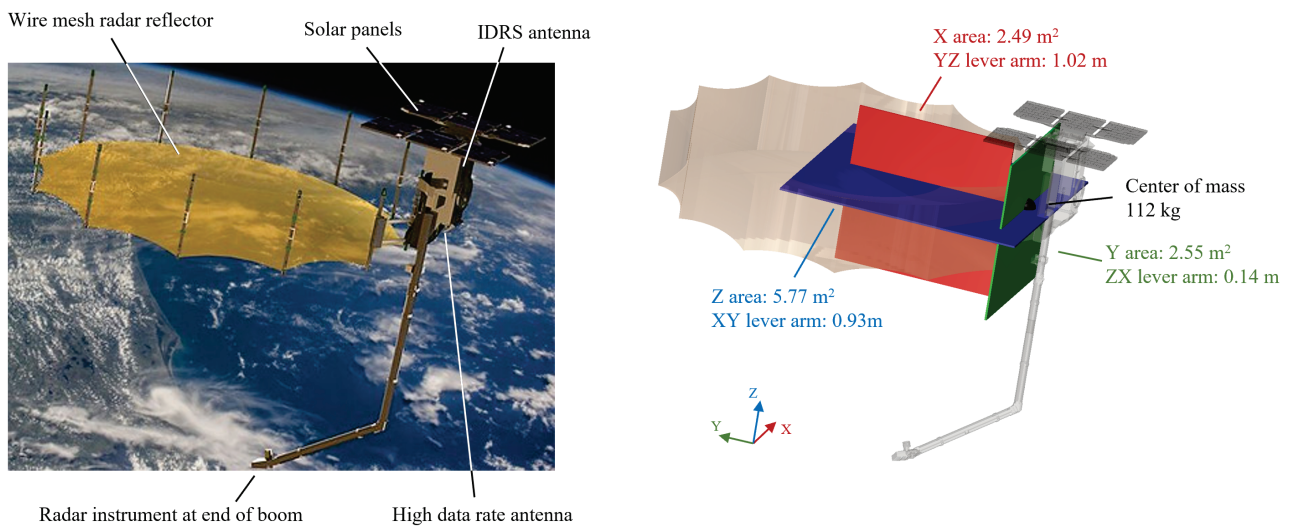

Figure 3: A rendering of a Whitney generation satellite in flight (left), and its 2022 flat plate aerodynamic model (right).

## 2.3 Capella's Satellite Concept of Operations (ConOps)

Since the Whitney satellite does not use any mechanically or electronically steerable components, the satellite must frequently change its orientation to point hardware components at different targets. Scheduled pointing activities include imaging collects, ground station contacts, and propulsive maneuvers. While time allocated to these activities for a given orbit varies significantly with different satellite schedules, over the long term these use approximately 30% of the satellite's time. The rest of the orbit is filled with background activities which depend on whether the satellite is in sunlight. When in sunlight, the satellite pivots to point its solar panels at the sun for power generation ("Sun pointing"). When in eclipse, the satellite pivots to point its IDRS antenna to track the nearest Inmarsat satellite to maintain a communication link ("IDRS pointing"). Over a time period of several days, analysis of telemetry has shown that the satellite's orientation to the velocity stream is well approximated by a uniform random sampling of the unit sphere.

## 2.4 Whitney Drag Estimation

The satellite's use of a large, high gain parabolic radar reflector is a key enabler of its best-in-class imagery data. However, even with the use of wire mesh reducing the area fill fraction of the reflector's geometric envelope (Figure 4), the reflector contributes a full three-quarters of the satellite's entire planar surface area at only one quarter of its mass.

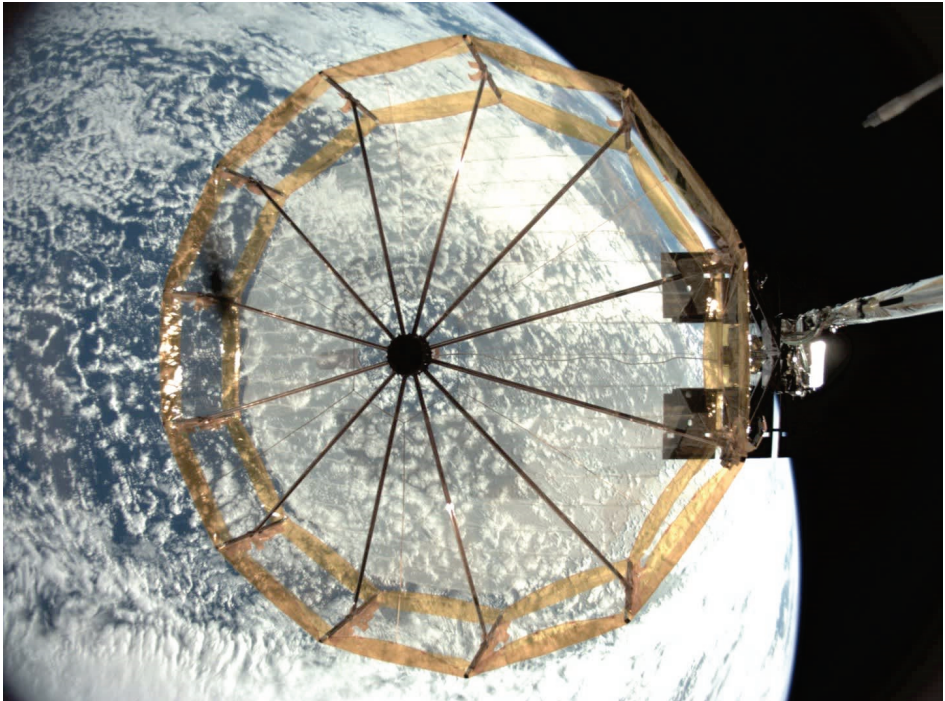

Figure 4: An on-orbit photograph of the fully deployed wire mesh radar reflector on the first Whitney-generation satellite Capella-3, taken from a camera at the end of its boom. The low fill fraction of the mesh gives the mesh optical transparency.

To estimate satellite ballistic coefficient prior to the launch of the first Whitney, Capella deemed that numerical simulation via Direct Simulation Monte Carlo (DSMC) would be unreliable, given the meshing difficulties of an enormous numbers of thin wires following a complex geometry. Instead, fill-fraction-weighted surface areas and centers of pressure were estimated and combined into a flat plate model for each of the primary spacecraft faces (Figure 3). Since the ConOps result in a random orientation of the spacecraft to the airstream, the overall effective area of the spacecraft was taken to be the average of these faces. During the early Whitney design effort in 2020 this was estimated at 3.0 m², and later revised in 2022 with better estimates to 3.6 m². The coefficient of drag was initially estimated at 2.2, using the default from NASA's Debris Analysis Software (DAS) version 3.1, which resulted in a 2020 ballistic coefficient estimate of $C_B = 0.059$ m²/kg.

During the Whitney design effort, Capella used this information along with the solar cycle prediction from the NOAA SWPC consensus model (Section 3.1) to predict atmospheric drag, which fed into an estimate of orbital lifetime. Analysis at that time found the propulsion system adequate to keep the spacecraft on-orbit for the 3-year design goal, by combining a propellant reserve for collision avoidance (COLA) maneuvers with drag compensation needs at a 525 km insertion altitude.

After launch, the ballistic coefficient was measured by fitting modeled satellite altitude decay using the NRLMSISE-00 density model [10] against measured orbit data, resulting in ballistic coefficient $C_B = 0.085$ m²/kg. From the satellite mass and updated area estimate, this implies a coefficient of drag of $C_d = 2.7$, which compares reasonably to estimates of shapes in near-vacuum molecular flow [11], but is also 44% higher than originally estimated.

A comparison to ballistic coefficients of other satellites is difficult to generate, as it depends on knowledge of satellite attitude and propulsive ConOps that is usually not publicly available. However, a literature review found a few collections of estimates for various satellites, shown below in Figure 5. From this, we can put in context Whitney's unique susceptibility to atmospheric drag as 2-10x higher than typical satellite missions.

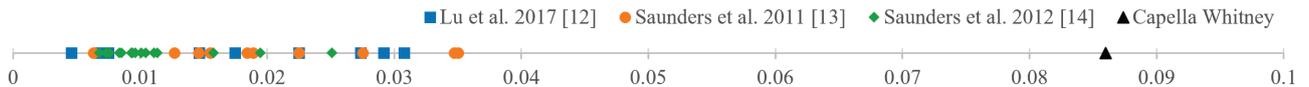

Figure 5: Estimates of ballistic coefficient $C_B$ for satellites from [12], [13], and [14], vs Capella's Whitney satellite. Note that the METEOR 1-1 and METEOR 1-7 $C_B$ estimates in [13] were hidden as they were not in family with the other METEOR satellites and believed to be erroneously high.

## 3     SOLAR CYCLE 25 and THE IMPACT on CAPELLA'S SATELLITES

### 3.1    Early Solar Cycle 25 Predictions

Ever since the end of Solar Cycle 21, the SWPC has convened a panel of experts during solar minimum to compare predictions for the upcoming solar cycle. The December 2019 prediction panel chaired by NOAA, NASA, and ISES published in December 2019 its "consensus" forecast of Solar Cycle 25, predicting a weak cycle in-line with Solar Cycle 24 [15] [16]. NASA Marshall Space Flight Center also releases a long-term prediction, updated monthly [17]. From 2020–2022, this model's mean prediction lined up with the SWPC panel, albeit with much larger variability. The other major long-term prediction of the cycle was put out in 2020 by the National Center for Atmospheric Research (NCAR). This used the newly hypothesized McIntosh et al. "Terminator" model of moving magnetic bands on the surface of the Sun, and contradicted the consensus to predict a stronger-than average solar cycle [18].

Common tools for analyzing satellite lifetime use different solar cycle predictions. Ansys' Systems Tool Kit's (STK's) satellite lifetime tool uses by default a prediction from its *SolFlx_CSSI.dat* file, last updated in 2017. The source of this prediction is unclear, but it aligns closely with the SWPC panel's model. STK's lifetime tool can also use its *space_weather.txt* data [19], which pulls from Celestrak Space Weather [20]. Celestrak prior to 26 May 2023 used the SWPC panel consensus prediction, after which it switched to using the NASA Marshall Solar Cycle Forecast. NASA's DAS tool uses the SWPC panel prediction for Solar Cycle 25 [21], however it appears to be biased towards the $-1\sigma$ value, presumably to bake in some conservatism to the tool for ensuring timely satellite reentry timelines. Table 1 shows the peak solar flux estimates from the major prediction sources, and Figure 6 plots the full predictions against the raw F10.7 solar flux observations from GFZ Helmholz Center Potsdam [22].

Table 1: Solar Cycle 25 Predictions. Predictions marked with * have been converted from sunspot number SSN to F10.7 solar flux via the 4th order model covering 1981-2015 from Table 9 in [1].

| Prediction Source | Prediction Date | Solar Cycle 25 Peak Flux | Uncertainty |
|---|---|---|---|
| NOAA SWPC prediction panel consensus [15] | 9 December 2019 | 138 SFU* in July 2025 | ± 8 SFU*, ± 8 months |
| Ansys STK lifetime tool: SolFlx_CSSI.dat [19] | January 2017 | 145 SFU in December 2025 | ± 9 SFU 95% confidence interval |
| NASA Marshall Solar Cycle Forecast [17] | August 2020 | 145 SFU in December 2023 | 102–229 SFU 90% confidence interval |
| | April 2022 | 138 SFU in December 2023 | 85–199 SFU 90% confidence interval |
| NASA DAS [5] [21] v3.2.0, v3.2.3 | 29 December 2021, 29 June 2022 | 135 SFU in July 2025 | - |
| McIntosh et al. [18] | 24 November 2020 | 221 SFU* in ~February 2024 | ± 10 SFU* 68% confidence interval, ± ~6 months |
| McIntosh/Leamon Unpublished 2022 Update [26] | 2 March 2022 | 194 SFU* in June 2025 | ± 13 SFU* 68% confidence interval |
| McIntosh/Leamon/Egeland Final Prediction [27] | 30 January 2023 | 190 SFU* | ± 12 SFU* 68% confidence interval |

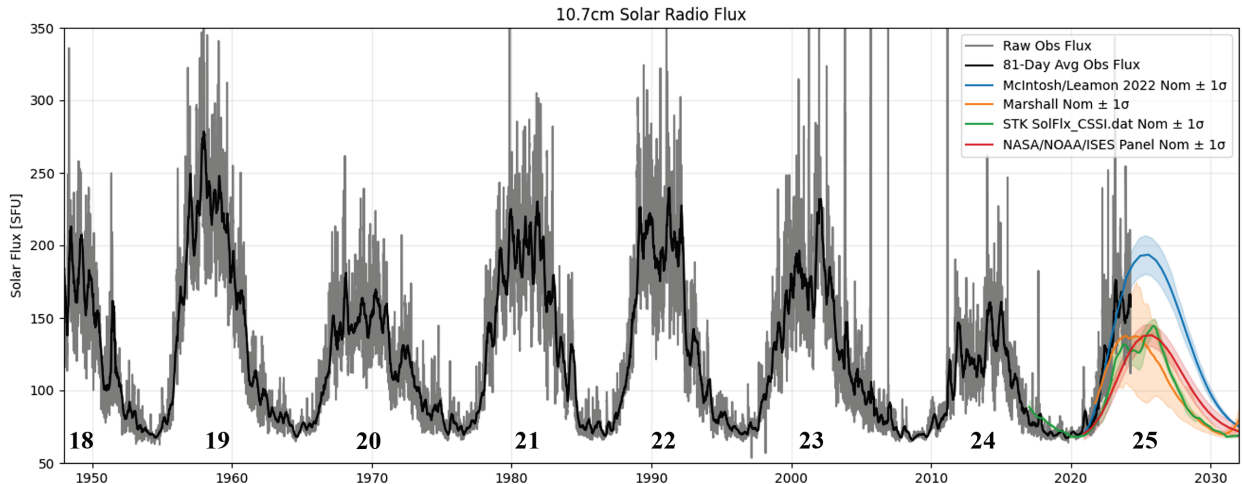

Figure 6: Historical observed F10.7 solar flux from [22], with predictions from Table 1 sources circa April 2022. Where applicable, confidence intervals were converted to $1\sigma$ values using a normality assumption. Numbers indicate the solar cycle.

While the stronger than expected solar cycle did not go unnoticed by the research community (see [23] from June 2022), the SWPC was slow to update its official estimates. It was not until October 2023 that it launched an experimental solar cycle prediction that updates monthly and better reflects the current state of the cycle [24].

### 3.2 Capella Experiences Increased Drag on Orbit

Capella first noticed an issue with atmospheric drag in April 2022, with noticeable acceleration of altitude decay across its constellation. For a representative 500 km circular orbit at $i = 45°$ inclination, the SWPC panel model predicted a density of 1.7e-13 kg/m$^3$, but the actual density over that month was around 4.2e-13 kg/m$^3$ (~2.5x higher), with peaks up to 9.9e-13 kg/m$^3$ (5.5x higher).

Searching for a better flux model, Capella found the NCAR team had recently announced that the Sun's "Terminator" event had occurred in December 2021, and that their internal forecasts had been updated based on this timing [25]. With the help of the Austrian Space Weather Office (ASWO) and its open-source "heliocats" tool [26], Capella was able to obtain a copy of this unpublished forecast and found it to match observations well. This 2022 McIntosh/Leamon model formed the basis of Capella's solar flux modeling moving forward, with a slight update after their finalized prediction in January 2023 [27]. In retrospect, the Marshall $1\sigma$-high prediction would also have worked.

To compare solar flux predictions, Capella chose a representative 500 km circular orbit at inclination $i = 45°$, and calculated the orbit-average density over all RAANs using the NRLMSISE-00 model at a density-weighted historical average geomagnetic Ap index of 10. This calculation was sped up by several orders of magnitude with the development of an internal tool that precalculated this data over a range of altitudes, inclinations, solar fluxes, and Ap indices, and then used 4-dimensional spline interpolation to find intermediate values. This density estimate later fed into another internally developed tool for probabilistic lifetime predictions. Figure 7 shows the differences in density predictions between the SWPC panel model and McIntosh/Leamon 2022 model for this representative orbit. Density over the whole solar cycle had increased by 2–3x, which decreases satellite lifetime a corresponding amount.

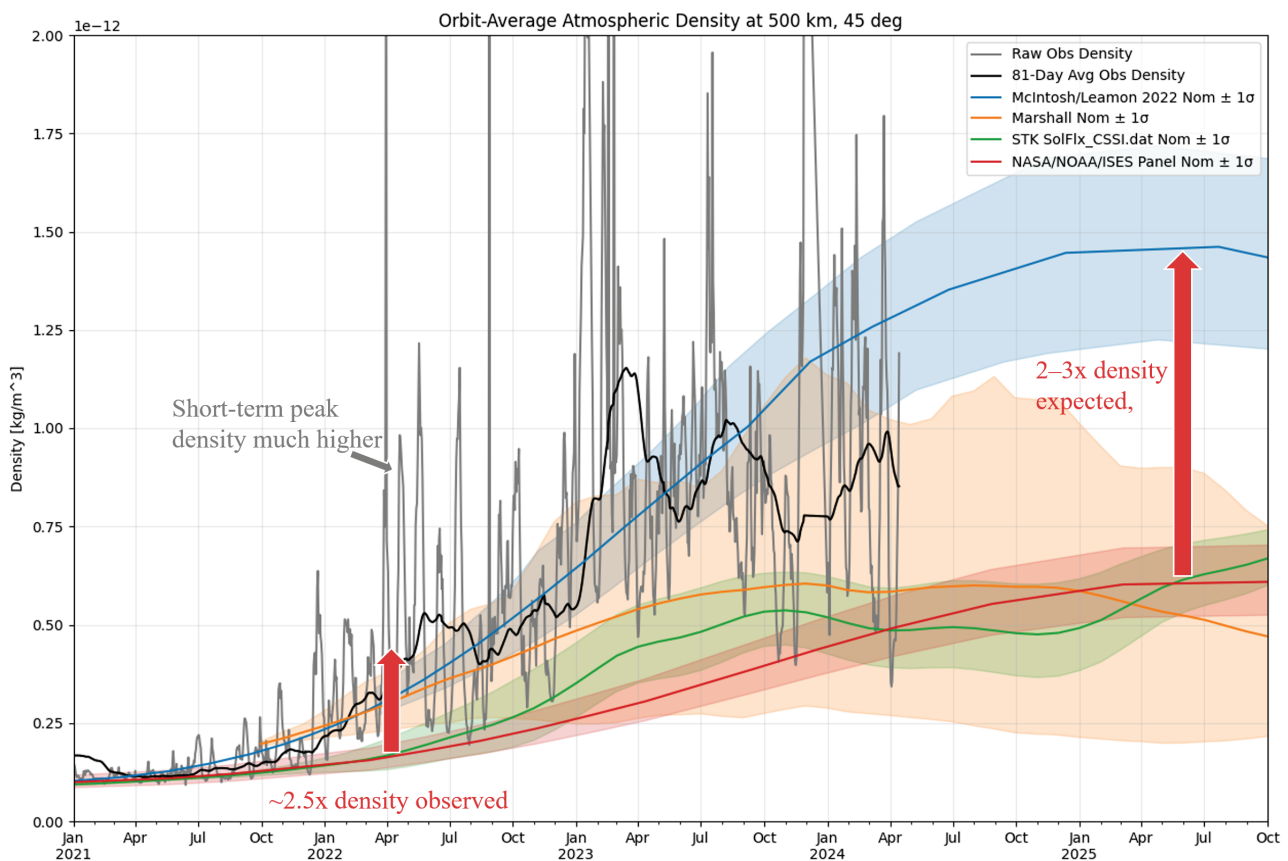

Figure 7: Density estimates for Solar Cycle 25 for a 500 km circular orbit at $i = 45°$ inclination, with predictions from Table 1 sources circa April 2022. Where applicable, confidence intervals were converted to $1\sigma$ values using a normality assumption.

This analysis predicted that if current trends continued, six of the seven spacecraft on-orbit would re-enter before the end of the year. Additionally, the upcoming Capella-9 spacecraft was slated for a launch to ~525 km, but at this insertion altitude would only survive for nine months even with full use of its onboard propellant. Mitigating the impending loss of these satellites instantly became a top priority for the company.

## 4 WAYS of INCREASING SATELLITE LIFETIME

Capella immediately began investigating strategies for increasing satellite lifetime. Different approaches broadly apply to different points in the satellite lifecycle:

1. Modifications to the baseline satellite design or mission architecture to make it less sensitive to drag. This was largely infeasible for Capella's situation, due to the extensive effort and time required to design and qualify a new satellite bus.
2. Subsystem modifications or upgrades, especially the propulsion system. While free space in Capella's internal bus volume is limited, its system architecture is modular and not tightly coupled. Subsystems can be "hot swapped", given adequate structural, power, and volume margins, and enough time for the associated design effort and procurement.
3. Choosing higher or more inclined orbits. For a small satellite operator flying on rideshare launches, this is largely constrained by launch provider orbit availability. However, using orbital transfer vehicles on a rideshare or procuring dedicated launches from "small launch" providers allows for more flexibility in orbit selection.
4. On-orbit ConOps modifications. This is the only way to mitigate drag for existing on-orbit assets.

### 4.1 Increase Altitude

Onboard propellant can be used to maintain altitude in the presence of atmospheric drag, or it can be used to perform orbit change maneuvers. Because atmospheric density continues to decrease as altitude increases, it is more propellant efficient from a lifetime perspective to raise the satellite as high as possible and coast back down than it is to hold altitude in a denser environment. Prior to the observation of increased on-orbit drag in 2022, Capella had performed a minimal number of propulsive maneuvers across the constellation in anticipation of future demonstrations of precision orbit control schemes. Lifetime concerns postponed these plans, in favor of immediate use of all propellant available for altitude raising across the entire constellation. Due to the low thrust of Whitney's electric thrusters and need to fit propulsive maneuvers between other activities in the busy satellite imaging schedule, these "burn campaigns" lasted several months. Figure 8 shows the effects of the successful burn campaigns on orbital altitude.

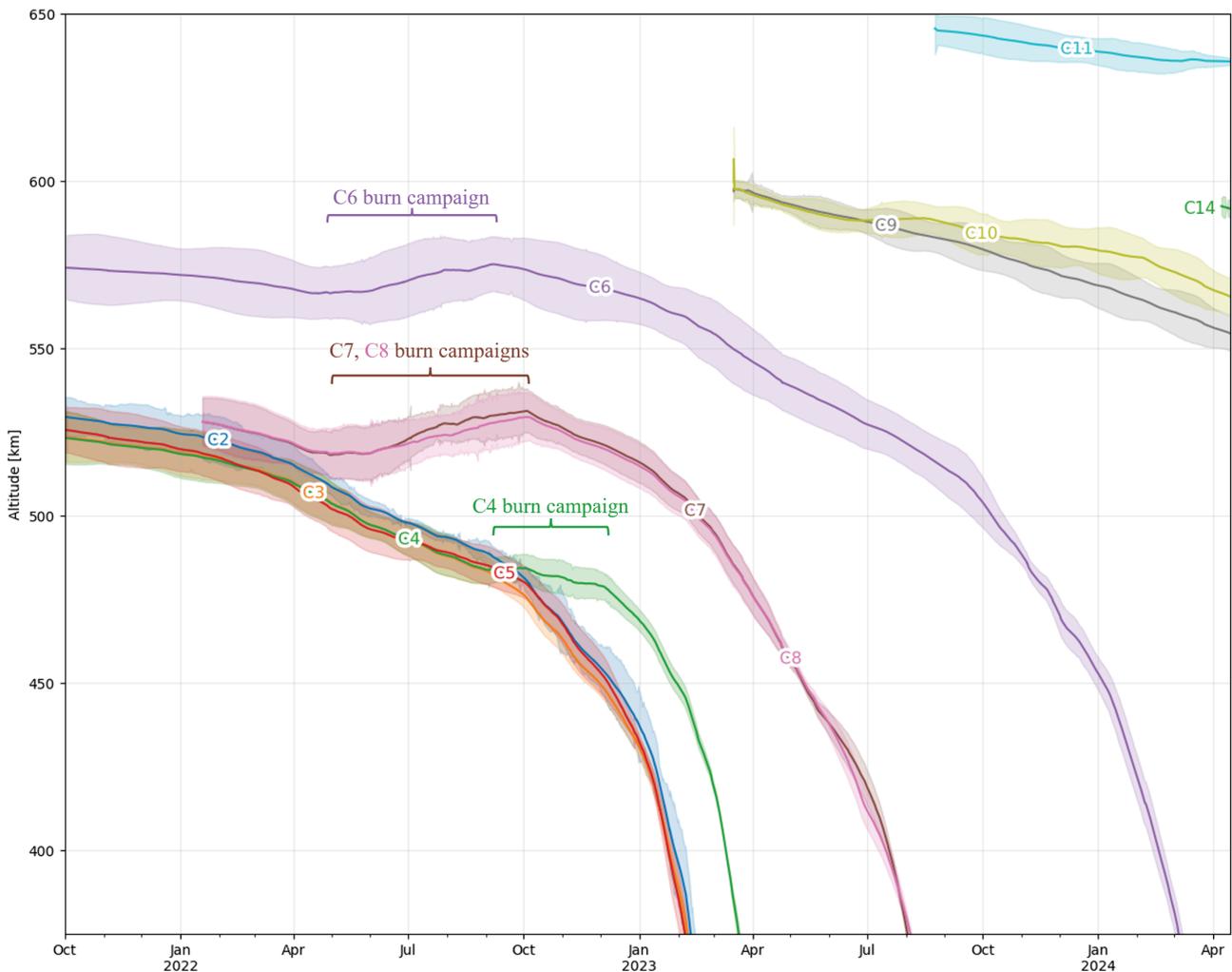

Figure 8: Orbital altitudes for Capella's satellite constellation over time, with burn campaigns labeled. The solid lines are semi-major axis altitude and the shading shows perigee to apogee.

Instead of burning propellant to raise altitude, it's far easier to simply be dropped off at a higher orbit by the launch vehicle. Being able to go higher than available rideshare missions was a major factor in Capella's decision to pull Capella-9 from its planned launch and procure a dedicated ride from Rocket Lab for Capella-9 and Capella-10, which were launched in March 2023 on an Electron rocket to a targeted 600 km [28]. This was also a major motivation behind the follow-on contract for four dedicated launches on Electron rockets for Acadia-generation satellites [29], the first of which launched to a targeted 640 km in August 2023.

There was a compromise to reach when choosing launch altitudes, between ensuring a minimum three-year operating life under conservatively denser solar flux profiles, and a maximum six year in-orbit life under conservatively less dense profiles for the streamlined small satellite FCC licensure. Meeting both requirements required Capella to reject the default solar flux profile in NASA's DAS tool, and substitute a custom profile based on the lower bound of the McIntosh model. See the Capella 9&10 Orbital Debris Assessment Report (ODAR) for details [30]. As a condition of accepting this rationale, Capella's FCC operating license stipulates that the company must continue to monitor space weather conditions, reserving propellant as needed to perform powered deorbits and not exceed an in-orbit lifetime of six years.

## 4.2 Increase Size of Propulsion System

Increased atmospheric drag has two implications for the sizing of a satellite propulsion system. First, the total impulse of the system must increase to counteract the cumulative drag which the satellite will experience over its lifetime. This can be accomplished with increased thruster efficiency (specific impulse) or more propellant mass. Second, given a power budget, thermal limitations on maximum burn duration, and ConOps scheduling constraints, thrust levels must be high enough to deliver adequate impulse *per orbit* to prevent altitude decay.

By mid-2022, Capella had already procured the propulsion systems for the last two Whitney satellites, as well as a planned propulsion upgrade for the first two Acadia-generation spacecraft Capella-11 and Capella-12, which increased total impulse by 5x. With the higher altitude afforded by dedicated Rocket Lab launches (Section 4.1), these systems were deemed adequate to meet a nominal three-year design lifetime even in elevated drag conditions.

However, thrust levels on this planned upgrade were deemed too low to deliver enough impulse per orbit for the elevated drag environment for future rideshare missions to more typical orbit insertion altitudes (500–550 km). So in late 2022, Capella kicked off design and procurement of a new electric propulsion system that had an additional 4x increase in thrust and 2.5x increase in total impulse. This new propulsion system flew and operated successfully for the first time on Capella-14 in April 2024.

## 4.3 Increase Orbital Inclination

At a given orbital radius, Earth's thermosphere is denser at the equator than at the poles due to more direct solar heating and the planet's equatorial bulge. Roughly, a satellite launched in an inclination $i = 0°$ equatorial orbit will encounter 30% higher orbit-averaged atmospheric density than one flying in an $i = 90°$ polar orbit (Figure 9). While raising altitude is a much more efficient use of onboard propellant than changing inclination after launch, inclination does play into orbit selection.

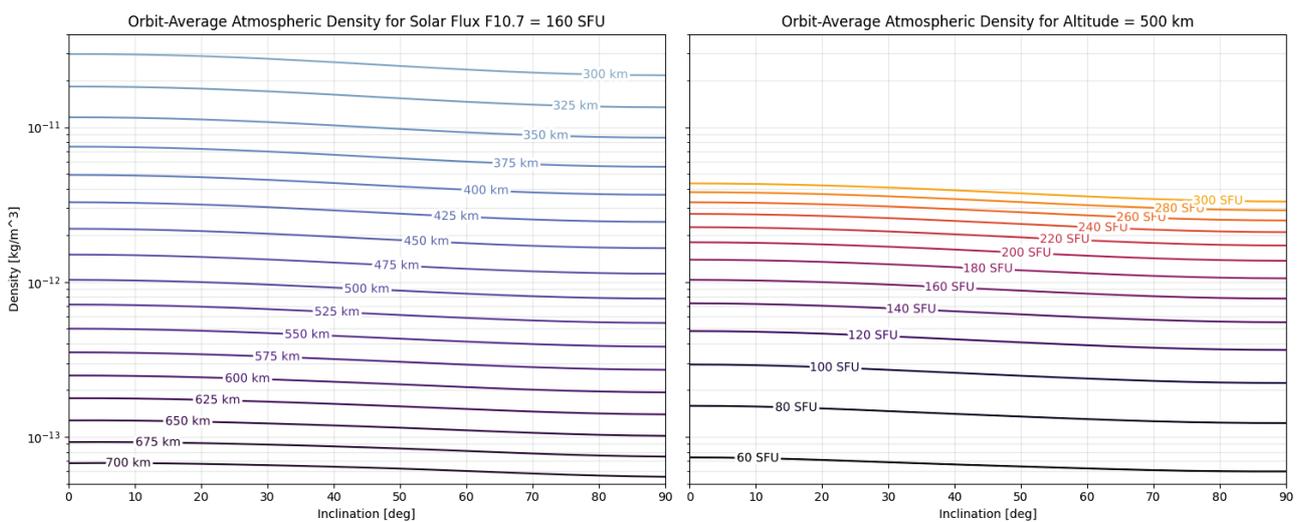

Figure 9: Orbit-averaged atmospheric density for a satellite flying in a circular orbit, as calculated using the NRLMSISE-00 density model and assuming a geomagnetic Ap index of 10 (see Section 3.2). Left shows density as a function of inclination and altitude during solar flux F10.7 = 160 SFU conditions, and right shows density as a function of inclination and solar flux at 500 km altitude.

## 4.4 Lower the Operational Altitude Floor

Capella's original plan for satellite operations was to capture images during orbital decay down to 450 km, but its ultimate floor for imaging operations is driven by the company's FCC license conditions and NOAA license issued by the Commercial Remote Sensing Regulatory Affairs office. Extending imaging operations to lower altitudes was an easy way to gain additional lifetime.

The ability of the satellite to operate at lower altitudes is also driven by its capacity to absorb aerodynamic torque transients and bleed off accumulated momentum with its attitude determination and control system (ADCS). Whitney's asymmetry and large aerodynamic lever arms (see Figure 3) stress its ADCS far more than a typical axisymmetric satellite design. To this end, Capella sized up the torque rods on all satellites past Capella-12 by 63%. For existing satellites on-orbit, Section 5.3 talks about ConOps adjustments that reduced accumulated momentum.

## 4.5 Increase Satellite Mass

Satellites are generally designed to be as mass efficient as possible, and the addition of raw mass is rarely desirable. But it does decrease the ballistic coefficient. Shortly before launch of Capella-11, Capella decided to add 10 kg of steel ballast mass to increase its mass to the upper end of the 160–170 kg range provided in its ODAR filing. This was estimated to add 3–4 months of life in a contingency scenario if its new propulsion system failed on first flight, and was left off later satellites.

## 4.6 Decrease Satellite Coefficient of Drag

Aerodynamically shaping the satellite to reduce its coefficient of drag was deemed of dubious value. Drawbacks were the complexity of performing iterative DSMC over a range of orientations, the lack of design heuristics for shapes in free molecular flow that reduce drag, the importance of the existing geometry to the radar imaging mission, and the likelihood that any benefits in coefficient of drag from added cowling would be more than negated from increases in overall surface area.

There have been a few proposals to decrease satellite coefficient of drag by smoothing surfaces, which increases the ratio of specular to diffuse molecular reflections and thus reduces skin friction. Extremely high levels of surface polishing [31] or atomic layer deposition of polymers [32] have been suggested as methods to accomplish this, however these surface treatments were deemed immature and in conflict with mesh reflector performance and bus thermal design.

## 4.7 Decrease Satellite Area

Removing or miniaturizing external components to reduce the satellite's area is only of benefit to decreasing drag if they are low enough mass to have a ballistic coefficient $C_B$ greater than the satellite average. No external parts on Whitney were identified as suitable to delete or substantially modify.

If the physical surface area of a satellite cannot be modified, then for a satellite with heterogenous geometry the *effective* surface area can be made smaller by modifying ConOps to fly while presenting smaller planar areas to the velocity stream. This is the principle of Capella's "low drag mode" flight.

## 5 LOW DRAG MODE

### 5.1 Low Drag Concept

By flying with its radar reflector and solar panels "edge-on" (see Figure 10), Whitney can avoid presenting its Z face to the velocity stream, which has ~2.2x the planar area of its X and Y faces (Figure 3). Of those two approximately equal-area options, the Y face is preferred for its lower aerodynamic lever arm, which is 14% that of the X face. Some intermediate orientation that perfectly balances aerodynamic torques could have been chosen, but with the uncertainties associated with the vehicle aerodynamics in free molecular flow, moving away from an orthogonal orientation of the spacecraft axes with respect to the airstream was deemed low impact and potentially confusing.

"Low drag mode" was thus defined as the Whitney satellite flying with its Y axis aligned with the velocity vector, with a rotation about that vector to best point the relevant hardware toward its target (e.g., solar panels towards the Sun). This pointing approach is also known as the "align-constrain" algorithm [33]. With Whitney's average planar area of 3.60 m$^2$ reduced to the Y area of 2.55 m$^2$ for the ~70% of flight spent conducting background activities, Capella expected a (schedule-dependent) ~21% reduction in drag, and approximately equal increase in satellite lifetime.

Changing a satellite's drag profile by adjusting its flight orientation is not a new concept. Several LEO satellite constellations have used differential drag as the primary method of orbit phasing and formation flying (see [34] section II for an overview), in lieu of an onboard propulsion system. Additionally, SpaceX has shared that their Starlink satellites were commanded into an edge-on low-drag orientation during the February 2022 geomagnetic storm [7], much like their regular "ducking" maneuvers to reduce frontal areas during conjunction events. However, the author is not familiar with this technique being adopted for active satellites that were not already designed for such a flight mode.

### 5.2 Trade-offs

The downside of flying in low drag mode is loss of precision pointing for power generation and IDRS communications link. IDRS connectivity was deemed easier to test than to model, but the impact to the power budget was more difficult to test since stressing conditions are not necessarily present at any given time in an orbit. So, power impacts were analyzed beforehand.

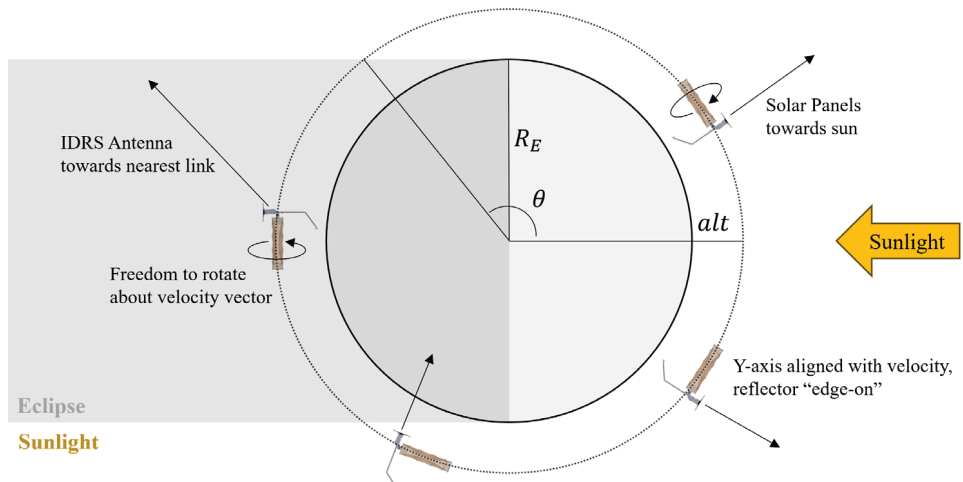

Figure 10: Low drag mode geometry when solar angle $\beta = 0°$.

The fraction of a circular orbit which is available for power generation depends on how much of that orbit is spent in eclipse. This is a complex calculation in the general case (see [35] and [36]). However, the only scenario of interest in this case was the minimum power stressing scenario. This happens when the solar $\beta$ angle (angle between the orbital plane and sun vector) is minimized at 0°, and the geometry of this scenario is shown in Figure 10. For a circular orbit in LEO where eclipse can be approximated as a cylinder rather than umbral and penumbral cones, the half angle of solar illumination $\theta$ is a function of the orbital altitude $alt$ and Earth's radius $R_E$ (Equation 3).

$$\theta = \frac{\pi}{2} + \cos^{-1}\left(\frac{R_E}{R_E + alt}\right) \qquad (3)$$

For baseline direct-pointing ConOps, the minimum-sunlight orbit that results in power generation fraction $p_{min}$ occurs at solar angle $\beta = 0°$ and is given by Equation 4.

$$p_{min} = \frac{1}{2\pi}\int_{-\theta}^{\theta} dx = \frac{\theta}{\pi} \qquad (4)$$

Solar angle $\beta = 0°$ is also the stressing case for the low drag mode align-constrain pointing algorithm. Since Whitney's solar panels point orthogonal to the Y axis, the optimally aligned solar pointing vector will be aligned with orbit radial or antiradial. The light hitting the solar panels will be attenuated by cosine losses, and so assuming no minimum incidence angle, the minimum power generation fraction in low drag mode is given by Equation 5.

$$p_{min} = \frac{1}{2\pi}\int_{-\theta}^{\theta} |\cos(x)| dx = \frac{2 - \sin(\theta)}{\pi} \qquad (5)$$

For a representative 500 km circular orbit, the ratio of Equations 4 and 5 show that low drag mode results in a 55% drop in power production capacity during the stressing low sunlight scenario. This was acceptable for the Whitney satellites that had completed their power-hungry burn campaigns.

**5.3  On-orbit Performance**

Capella-7 and Capella-8 were launched together on the SpaceX Transporter-3 mission in January 2022, and after their burn campaign their orbital altitudes gradually decayed until they re-entered the atmosphere in August 2023. Since they shared an orbit, they provided an excellent natural experiment to compare the effects of low drag mode. In May of 2023, low drag mode was enabled on Capella-7, with Capella-8 remaining in its baseline ConOps as a reference. After a month, the flight modes were swapped to confirm the generality of the results, and this configuration was kept through reentry.

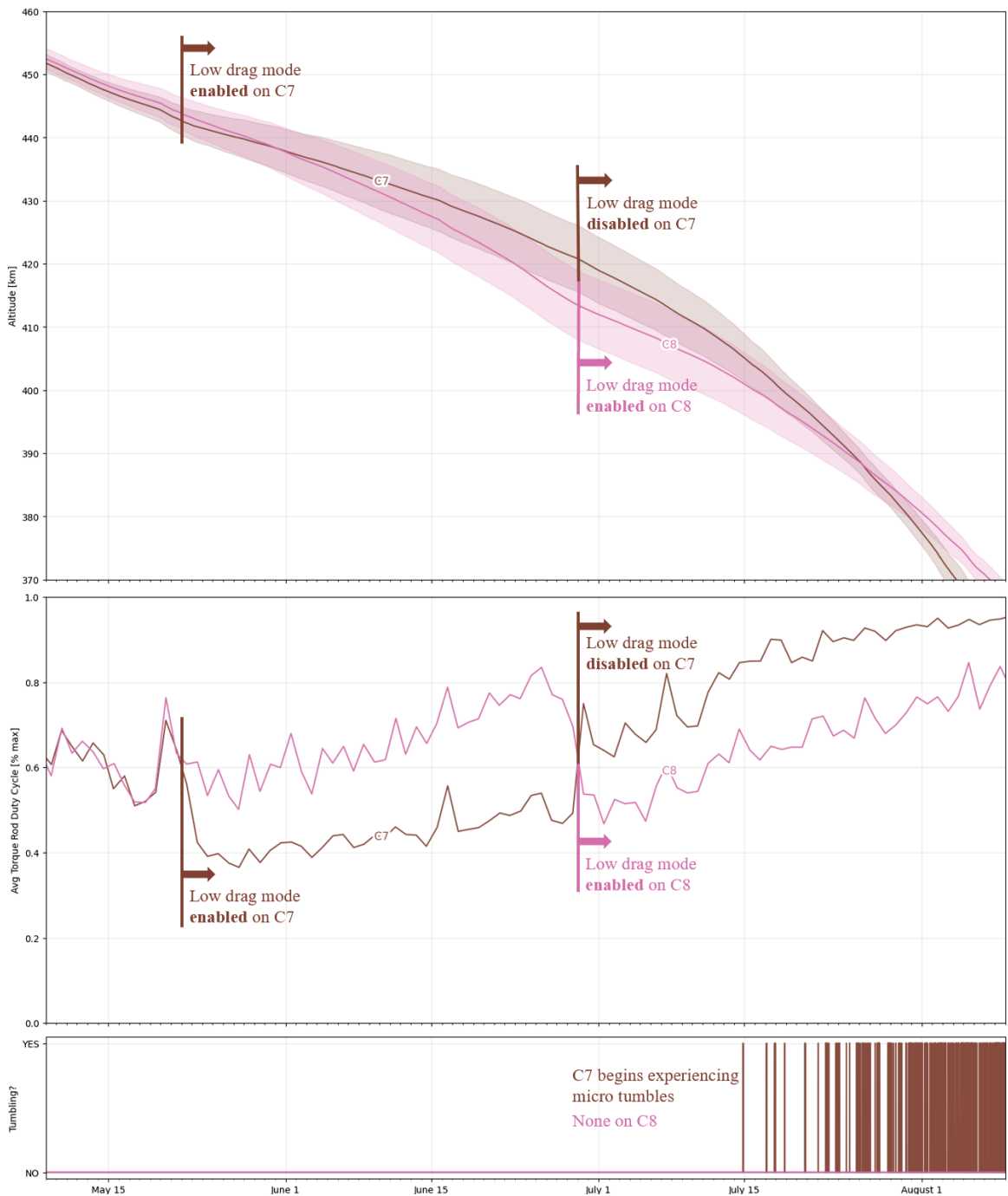

Figure 11: The effects of low drag mode on Capella-7 and Capella-8. Top shows altitude over time, where the solid lines are semi-major axis altitude and the shading shows perigee to apogee. Middle is daily average torque rod duty cycle, which tracks momentum accumulation due to aerodynamic torques. Bottom shows micro tumble events.

Figure 11 shows the effects of low drag mode on satellite altitude decay, with the change in behavior being visually apparent in plots of altitude over time. By fitting predicted decay against actuals, Capella found that low drag mode resulted in a 24% decrease in aerodynamic drag for both spacecraft, comparing well to the ~21% predicted.

Figure 11 also shows the effect of low drag mode on torque rod duty cycles, which is taken as a proxy for accumulated spacecraft momentum from aerodynamic torques. The satellite with low drag mode enabled consistently showed 20–30% lower torque rod duty cycles. Additionally, towards the end of a spacecraft's life as it falls to denser layers of the atmosphere, it begins experiencing "micro tumble" events where the reaction wheels saturate and are unable to control attitude for several minutes. Low drag mode lowered the altitude where these started occurring from 404 km on Capella-7, to 365 km on Capella-8.

The downsides of reduced IDRS link and power accumulation were assessed during this test. While high-fidelity power model validation was not performed due to limitations in telemetered data, the satellites in low drag mode maintained adequate battery charge levels. IDRS link was characterized by the duration of outage between subsequent connections. Figure 12 shows a 35% increase in mean outage duration and longer tail cases, but Capella found this to have minimal impact on operations.

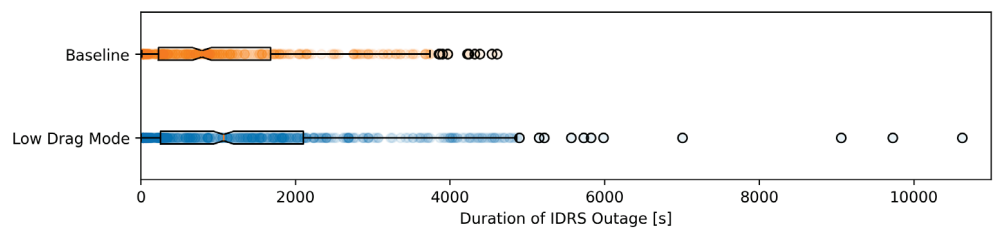

Figure 12: Boxplot of the effects of low drag mode on IDRS outage duration.

One more unexpected benefit of low drag mode was observed. Reducing the susceptibility to drag on the spacecraft by ~24% also reduced the *variability* of medium-term orbit propagation predictions. This improved the timing accuracy when scheduling imaging collects and ground downlink contacts.

## 6  LESSONS LEARNED

There are several lessons to be learned from Capella's experience:

- Solar physics is still a nascent field, and solar cycle predictions have significant epistemic uncertainty outside of the aleatory uncertainty within each prediction. This extends not only to the magnitude and timing of the solar cycle, but to whether there will be a solar cycle *at all* — we do not yet have an explanation for the "Maunder Minimum" between 1645–1715 where solar activity remained at prolonged low levels. For Solar Cycle 26, Capella recommends sizing spacecraft propulsion systems based on historical extremes regardless of the forecast.

- The default solar cycle prediction or coefficient of drag in common satellite lifetime tools should not be blindly trusted for satellite lifetime design purposes.

- Propulsion is a system worth oversizing. Deorbit from LEO is a hard lifetime limiter, and the added value from marginal mission extension far exceeds the cost of additional propellant. Solid propellant electric thrusters currently in development will make propulsion a much more compact system to scale up by the time Solar Cycle 26 arrives.

- Differential drag or low drag flight modes are feasible ConOps even for vehicles not specifically designed for them, though they come with power and other operational trade-offs.

- Satellite lifetime estimation needs to be a continuously monitored metric after launch, rather than being just a design requirement. These estimates should be driven by real-time telemetry and the latest solar cycle prediction updates. They should also be a probabilistic metric for the range of possible futures rather than a point estimate.

- Vertical integration within a satellite company of design, manufacture, and operations is extremely powerful. Capella was able to quickly tackle its drag problem at all levels of the satellite lifecycle, and trade off operational solutions against hardware solutions with clear visibility into both sides.

- This is a case study that proves out the value of dedicated small launch vehicles to provide the flexibility and agility that satellite operators may need in unique situations.

- This is also a case study to show the value of a proliferated LEO constellation architecture with regular launches and iterative improvement. Satellites lost to drag were able to be backfilled by satellites already in production, and the designs of the satellites in production were able to be modified in real time to better handle the changing space environment.

# 7 SUMMARY

In April of 2022, Capella Space discovered that their Whitney satellites were deorbiting faster than NOAA SWPC's consensus solar cycle model predicted, due to increased solar flux densifying the atmosphere in LEO to 2–3x over predictions. Based on on-orbit data that increased Capella's estimate of Whitney's ballistic coefficient from 0.059 to 0.085 m$^2$/kg and the 2022 McIntosh/Leamon solar flux model, Capella revised their satellite lifetime estimates and found that they expected to lose six out of their seven operational satellites by the end of the year.

Faced with the impending early loss of its satellites, Capella acted quickly and decisively to address the increased drag environment and its effects. The company immediately started burn campaigns to raise orbits of its on-orbit satellites, procured a new more performant electric propulsion system, switched launch providers for upcoming missions to "small launch" dedicated rockets that could reach higher orbits, added ballast mass to already-built satellites to extend life, sized up torque rods on future satellites to lower the operational altitude floor, and deployed and validated a new "low drag" flight mode that extended in-orbit lifetime by 24%. While all the satellites that were flying in April 2022 have since reentered, these efforts extended the lives of those vehicles by a cumulative 2.5 years, and that of Capella's expanding constellation by many times that.

# 8 ACKNOWLEDGEMENTS

The author thanks the entire Capella Space team for its rapid response and hard work solving the satellite drag situation. Thanks also to Scott McIntosh of NCAR and Chris Möstl of ASWO for their assistance in obtaining and interpreting pre-released solar cycle predictions.